\documentclass[a4paper,fleqn,usenatbib]{mnras}
\usepackage{amsmath}
\usepackage{epsfig} 
\usepackage{graphicx}	
\usepackage{amssymb}	

\newcommand{\be}{\begin{equation}}
\newcommand{\e}{\end{equation}}
\newcommand{\bear}{\begin{eqnarray}}
\newcommand{\ear}{\end{eqnarray}}

\newcommand{\del}{\partial}

\def\aj{AJ}
\def\apj{ApJ}
\def\apjs{ApJS}
\def\jcap{JCAP}
\def\mnras{MNRAS}
\def\aap{A\&A}
\def\prd{Physical Review D}

\def\nat{Nature}      
\def\apjs{ApJS}
\def\apjl{ApJ Letters}
\def\physrep {Physics Reports}

\title[Entropy of the Cosmic Web] {Configuration entropy of the Cosmic
  Web: Can voids mimic the dark energy?}
    
\author[Pandey, B.] {Biswajit Pandey\thanks{E-mail:
    biswap@visva-bharati.ac.in}
  \\ Department of Physics, Visva-Bharati University, Santiniketan,
  Birbhum, 731235, India\\ }
   
 \date{\today}

 \pubyear{2017}
  
\begin{document}
\label{firstpage}
\pagerange{\pageref{firstpage}--\pageref{lastpage}}      
\maketitle
       
\begin{abstract}
We propose an alternative physical mechanism to explain the observed
accelerated expansion of the Universe based on the configuration
entropy of the cosmic web and its evolution. We show that the sheets,
filaments and clusters in the cosmic web act as sinks whereas the
voids act as the sources of information. The differential entropy of
the cosmic velocity field increases with time and also act as a source
of entropy. The growth of non-linear structures and the emergence of
the cosmic web may lead to a situation where the overall dissipation
rate of information at the sinks are about to dominate the generation
rate of information from the sources. Consequently, the Universe
either requires a dispersal of the overdense non-linear structures or
an accelerated expansion of the underdense voids to prevent a
violation of the second law of thermodynamics. The dispersal of the
sheets, filaments and clusters are not a viable option due to the
attractive nature of gravity but the repulsive and outward peculiar
gravitational acceleration at the voids makes it easier to stretch
them at an accelerated rate. We argue that this accelerated expansion
of the voids inside the cosmic web may mimic the behaviour of dark
energy.

\end{abstract}


\section{Introduction}

The current accelerated expansion of the Universe remains one of the
major unsolved puzzle in cosmology. Observations imply that we live in
an expanding Universe \citep{hubble} which is currently expanding at
an accelerating rate \citep{riess, perlmutter}. The accelerated
expansion is unforeseen in an Universe in the presence of matter and
gravity. The observed acceleration can be explained by the general
theory of relativity only if a hypothetical entity called dark energy
with a negative pressure co-exists along with the matter and radiation
in the Universe.

Einstein introduced the cosmological constant $\Lambda$ in his field
equations to counterbalance the effect of gravity and achieve a
stationary Universe. But such a time-independent cosmological constant
can be also regarded as a candidate for dark energy. The cosmological
constant is very often identified as the energy of the vacuum which
remains constant despite the expansion of the Universe and eventually
becomes the most dominant component leading to the observed
acceleration. Unfortunately, the theoretical value of vacuum energy
predicted by the quantum field theories exceeds the observed value of
the cosmological constant by a factor of $10^{60}$ to $10^{120}$. The
huge discrepancy between the predicted theoretical value and the
observational value indicates that the nature and physical origin of
the dark energy still remain largely elusive.

 Many alternative models of dark energy have been proposed by either
 modifying the matter side (e.g. quintessence \citep{ratra, caldwell}
 and k-essence \citep{armendariz}) or the geometric side (e.g. f(R)
 gravity \citep{buchdahl} and scalar tensor theories
 \citep{bransdicke}) of Einstein's field equations. Some other
 alternatives have been also introduced based on some interesting
 physically motivated ideas such as the backreaction mechanism
 \citep{buchert2k}, effect of a large local void \citep{tomita01,
   hunt}, entropic force \citep{easson}, extra-dimesnion
 \citep{milton}, entropy maximization \citep{radicella, pavon1},
 information storage in the spacetime \citep{paddy, paddyhamsa} and
 configuration entropy of the Universe
 \citep{pandey1}. \citet{copeland} and \citet{de2010} provide a
 detailed review on some of these models and the possible ways to
 confront them with observations.
 
The $\Lambda$CDM model with a time-independent cosmological constant
still stands out as the best bet among all the possible scenarios
proposed in the literature. The cold dark matter (CDM) model was put
forth by \citet{peebles} in the early 1980s. The CDM model very soon
became the central ingredient in understanding the cosmic structure
formation \citep{bond,blue,davis}. Subsequent observations constrained
the total matter density of the Universe to $\Omega_{m}\sim 0.3$
\citep{carlberg, mohr} and the observations of the CMBR suggested a
critical density universe with a flat FRW geometry \citep{komatsu,
  planck}. This leaves us with the only choice that
$\Omega_\Lambda=0.7$. Though the $\Lambda$CDM model currently explains
most of the observations but we hardly understand the $\Lambda$ in
this model.

The CMBR observations suggest that the Universe was highly smooth and
regular at earlier times. The level of anisotropy observed in the CMBR
is $\sim 10^{-5}$. On the other hand, the present day mass
distribution in the Universe is highly irregular and clumpy due to the
structure formation. \citet{pandey1} proposed an alternative scenario
where the dissipation of the configuration entropy in the Universe due
to structure formation can lead to an accelerated expansion of the
Universe. If the other entropy generation mechanisms are not as
efficient as the dissipation then the Universe requires a mechanism to
cease the dissipation by suppressing the growth of
structures. Interestingly, introducing a cosmological constant
$\Lambda$ in the setting shut off the growth of structures on large
scales and terminate the dissipation of the configuration entropy.
However, this is not true for the non-linear structures for which the
collapse has already started. The anisotropic gravitational collapse
\citep{zeldovich} leads to the formation of a complex network known as
the ``Cosmic Web'' \citep{bond1}. The earlier analysis by
\citet{pandey1} was limited to the linear regime. In the present work,
we extend the analysis beyond the linear regime using the Zeldovich
approximation \citep{zeldovich} and propose a possible mechanism for
the cosmic acceleration driven by the information flow inside the
cosmic web.

\section{CONFIGURATION ENTROPY OF THE COSMIC WEB AND ITS EVOLUTION}
We consider a large volume $V$ of the Universe and treat its matter
content as a fluid. The configuration entropy of the fluid in that
volume is then defined \citep{pandey1} as,
\begin{eqnarray}
S_{c}(t) & = & -\int \rho(\vec{x},t)\log \rho(\vec{x},t)\, dV
\label{eq:one}
\end{eqnarray}
where $dV$ is a small element of volume and $\rho(\vec{x},t)$ is the
density inside the volume element centered at $\vec{x}$. The
definition is motivated by the idea of the information entropy
originally proposed by \citet{shannon48}.

\citet{pandey1} show that the configuration entropy in a volume $V$
of the fluid will evolve as,
\begin{eqnarray}
\frac{dS_{c}}{dt} & = & \int \rho(\vec{x},t) \, \nabla \cdot \vec{v}
(\vec{x},t) \, dV
\label{eq:two}
\end{eqnarray}
where $\vec{v}(\vec{x},t)$ is the peculiar velocity of the fluid
contained inside the volume element dV at time t.

Zeldovich approximation \citep{zeldovich}, a first order
Lagrangian perturbation theory provides an elegant description of the
formation of the cosmic web. It provides a scheme for mapping the
Lagrangian positions of the particles to their Eulerian
co-ordinates. If $\vec{x}(t)$ is the Eulerian co-ordinate of a
particle at time t then it is related to its initial Lagrangian
co-ordinate $\vec{q}$ at $t\rightarrow 0$ as,
\begin{eqnarray}
\vec{x}(t)=a(t)[\vec{q}+D(t)\,\vec{S}(\vec{q})]
\label{eq:three}
\end{eqnarray}
where $a(t)$ is the scale factor, $D(t)$ is the growing mode of
density perturbations and $\vec{S}(\vec{q})$ is a time-independent
vector field. It assumes that the particles continue to move along the
initial directions.

Considering the conservation of mass, one can obtain the Eulerian
density $\rho(\vec{x},t)$ as,
\begin{eqnarray}
\rho(\vec{x},t)=\frac{\bar{\rho}}{|\frac{\del x_{i}}{\del q_{j}}|}
\label{eq:four}
\end{eqnarray}
where $\bar{\rho}$ is the mean density at time t and $\frac{\del
  x_{i}}{\del q_{j}}$ is the Jacobian of the transformation between
$\vec{x}(t)$ and $\vec{q}$. This is often known as the deformation tensor
which accounts for the gravitational evolution of the fluid. The
vector field $\vec{S}(\vec{q})$ is irrotational since the density
perturbations originate from the growing mode. This allows one to
write it in terms of a potential. Consequently, the deformation tensor
becomes a real symmetric matrix which is diagonalizable in some
co-ordinate system. The Eulerian density can be written as,
\begin{eqnarray}
\rho(\vec{x},t)=\frac{\bar{\rho}}{(1-D(t)\lambda_{1}(q))(1-D(t)\lambda_{2}(q))(1-D(t)\lambda_{3}(q))}
\label{eq:five}
\end{eqnarray}
where $\lambda_{i}=\frac{\del x_{i}}{\del q_{i}}$ are the eigenvalues
of the deformation tensor. $\lambda_{1}(q)$, $\lambda_{2}(q)$ and
$\lambda_{3}(q)$ are the three eigenvalues which give contraction or
expansion along the three eigenvectors. If the eigenvalues are ordered
as $\lambda_{1}(q)>\lambda_{2}(q)>\lambda_{3}(q)$ then the
gravitational collapse would first occur along the shortest axis
corresponding to the largest eigenvalue. The first singularity occurs
at $q$ when $D(t)=\frac{1}{\lambda_{1}(q)}$. The contraction along
the shortest principal axes leads to a sheetlike structure which are
believed to be the first non-linear structure formed by gravitational
clustering. \citet{doros} show that simultaneous collapse along
multiple axes is unlikely to occur. The subsequent collapse along the
medium and the longest principal axes would produce a filament and a
cluster respectively.

The Zeldovich approximation predicts the emergence of the observed
cosmic web \citep{doros80,pauls,sarkar18} and provides a fairly good
match to the structures predicted by the N-body simulations
\citep{buchert89,coles93,yoshisato,tassev}. Some reviews of the
Zeldovich approximation can be found in \citet{shandarin89},
\citet{sahni95}, \citet{hidding14} and \citet{white14}.

The sheets, filaments and clusters in the cosmic web would emerge at a
given location depending on the signs of the eigenvalues
$\lambda_1(q)$, $\lambda_2(q)$ and $\lambda_3(q)$. When one of the
eigenvalues is positive and the other two are negative then the
anisotropic gravitational collapse would produce a sheetlike
structure. Similarly two positive and one negative eigenvalues would
produce a filament and all three positive eigenvalues would
eventually produce a cluster. In all of the above three cases the
density $\rho(\vec{x},t)$ at a location $\vec{x}$ increases with time
eventually reaching singularity when collapse occur along one or
multiple eigenvectors. On the other hand, if all three eigenvalues of
the deformation tensor are negative then the density at a location
$\vec{x}$ would continuously decrease producing a large underdensity
or void. 

The divergence of the peculiar velocity field $\nabla \cdot \vec{v}$
would be negative at the locations where sheets, filaments or clusters
are formed. This is simply due to the inflow of mass towards these
overdensities from their surroundings. The nature of flow would be
different around these structures but the sign of $\nabla \cdot
\vec{v}$ would be always negative for these structural elements of the
cosmic web. On the other hand, the mass outflow from the
underdensities or the voids towards the neighboring overdensities at
their periphery would always give rise to a positive $\nabla \cdot
\vec{v}$. Combining these information with \autoref{eq:two} indicates
that the configuration entropy $S_c(t)$ would always dissipate from
the overdense regions whereas it would increase inside the underdense
regions. The decrease and increase in the configuration entropy in
different parts of the cosmic web could play a crucial role in
governing the dynamics of the cosmic web.

One should also take into account the change in the information
entropy of the cosmic velocity field besides the change in the
configuration entropy of the mass distribution inside the cosmic web.
It is difficult to analytically predict the detailed flow patterns
around the sheets, filaments and clusters in the cosmic web. It would
require N-body simulations to fully track the details of the cosmic
velocity field. Even with the N-body simulations, the collisionless
dark matter makes the task complicated due to the multi-valued nature
of the velocity field after shell crossing and discontinuities at
caustics \citep{hahn}.

We consider a peculiar velocity field $\vec{v}(\vec{x})$ which is
smoothed over a radius $R$ using a spherical top hat or a Gaussian
window function of size R. The smoothed peculiar velocity field is
given by,
\begin{eqnarray}
\vec{v}_{R}(\vec{x})= \int \vec{v}(\vec{x})
W_{R}(\vec{x}-\vec{x^{\prime}})\, d^{3}x^{\prime}
\label{eq:six}
\end{eqnarray}
where $W_{R}(\vec{x}-\vec{x^{\prime}})$ is the window function used
for smoothing. The radius R is chosen such that $R<<L$, where
$L=(\frac{3V}{4\pi})^{\frac{1}{3}}$ and V is the volume considered.

The line of sight component of this smoothed peculiar velocity field
is $v_{R}(\vec{x})=\vec{v}_{R}(\vec{x})\cdot
\frac{\vec{x}}{|\vec{x}|}$. The dispersion in $v_{R}(\vec{x})$ can be
written as \citep{seto},
\begin{eqnarray}
X(L,R)= \frac{1}{V}\int v_{R}^2(\vec{x}) \, d^{3}x.
\label{eq:seven}
\end{eqnarray}

The normalized dispersion of the line of sight component of the
peculiar velocity field is given by,
\begin{eqnarray}
\sigma_{R}^2= \frac{\langle X^{2}(L,R)\rangle-\langle
  X(L,R)\rangle^{2}}{\langle X(L,R)\rangle^{2}}.
\label{eq:a_seven}
\end{eqnarray}

The cosmic velocity field can be described by a Gaussian distribution
in the mildly non-linear regime \citep{nusser,ciecielag}. The
differential entropy of a Gaussian distribution is $\ln(\sigma
\sqrt{2\pi e})$ where $\sigma$ is the standard deviation of the
Gaussian. In the present context, if the distribution of the line of
sight component of the peculiar velocity can be described by a
Gaussian then the entropy associated with it would be $\ln(\sigma_{R}
\sqrt{2\pi e})$, where $\sigma_{R}$ is given by
\autoref{eq:a_seven}. The value of $\sigma_{R}$ increases with the
growth of non-linear structures resulting in an increase in the
entropy associated with the velocity field.

It would be also interesting if the evolution of the configuration
entropy can be directly related to the deceleration parameter. The
deceleration parameter $q=-\frac{\ddot{a}a}{\dot{a}^2}$ is a
dimensionless measure of the acceleration or deceleration of the
expansion of the Universe. It varies with time and the nature of
variation depends on the background cosmological model. Observations
suggest that its present value $q_0$ is negative which indicates that
the Universe is currently undergoing an accelerated expansion. 

The evolution of the configuration entropy $S_c$ inside a large
comoving volume V can be described by the following equation
\citep{pandey1,das},
\begin {eqnarray}
  \frac {dS_c(a)}{da} + \frac {3}{a}(S_c(a) - M) + \bar\rho f(a) \frac
        {D^2(a)}{a}\int \delta^2(\vec x)\, dV = 0
  \label {eq:eight}
\end {eqnarray}
where, a is the scale factor, $D(a)$ is the growing mode of density
perturbation and $f(a)= \frac {dlnD}{dlna} = \frac {a}{D} \frac
{dD}{da}$ is the dimensionless linear growth rate.

Initially, the growth of density perturbations are negligible and the
evolution of the configuration entropy is decided by the initial
condition. When the contribution of the third term in
\autoref{eq:eight} is ignored, the analytical solution of
\autoref{eq:eight} is given by,
\begin {eqnarray}
  \frac {S_c(a)}{S_c(a_i)} =
  \frac{M}{S_c(a_i)}+\Bigg(1-\frac{M}{S_c(a_i)}\Bigg)\Bigg(\frac{a_i}{a}\Bigg)^3
  \label {eq:nine}
\end {eqnarray}
where $a_i$ is the initial scale factor. We expect an initial
transient in the configuration entropy when $S_c(a_i)>M$ or
$S_c(a_i)<M$. No such transient behaviour is expected when
$S_c(a_i)=M$. The cosmology dependence of the evolution comes into
play only after significant growth of structures has taken place. The
evolution of the configuration entropy is then controlled by the third
term in \autoref{eq:eight}. Here, we consider $S_c(a_i)=M$ as we are
only concerned with the cosmology dependence. We set the time
independent quantities in the third term of \autoref{eq:eight} to 1 for
simplicity. Defining $P(a)=f(a) D^2(a)$, the \autoref{eq:eight} can be
written as,
\begin {eqnarray}
  \frac {dS_c}{dt} = - \frac {\dot{a}}{a} P(a).
  \label {eq:ten}
\end {eqnarray}

Differentiating \autoref{eq:ten} with respect to time and rearranging,
the deceleration parameter can be expressed as,
\begin {eqnarray}
  q=\frac{1}{H^2(a) P(a)}\frac{d^2S_c}{dt^2}+\frac{a}{P(a)} \frac{dP(a)}{da}-1.
  \label {eq:eleven}
\end {eqnarray}

$H(a)$, $P(a)$ and $\frac{dP(a)}{da}$ can be calculated for any given
cosmological model.  $\frac{d^2S_c}{dt^2}$ in \autoref{eq:eleven} is
the time derivative of the entropy rate shown in \autoref{eq:two}. So
we can calculate q at any given epoch by measuring the derivative of
the entropy rate and using it in \autoref{eq:eleven}. However, it is
difficult to calculate the entropy rate and its derivative in the
mildly non-linear regime without the help of N-body simulations.

\section{DISCUSSION AND CONJECTURE}
Let us imagine a sufficiently large volume V over which the Universe
can be treated as homogeneous and isotropic. If we consider the
Universe to be divided into many such volumes then there will be no
net mass exchange across these volumes. Initially the mass
distribution would be highly uniform in each of these volumes leading
to a high configuration entropy for the mass distribution. But over
time, the tiny density fluctuations present in them would be amplified
by gravitational instability leading to the formation of the cosmic
web. The segregation of mass into different morphological components
such as sheets, filaments and clusters would produce a highly
non-uniform distribution inside each of these volumes. The mass which
was earlier distributed uniformly across the entire volume V now only
occupies a small fraction of it and is distributed in a complex
filamentary cosmic web. This leads to an overall decrease in the
configuration entropy of the mass distribution inside the volume
V. The \autoref{eq:five} tells us that the density would increase at
the regions where sheets, filaments and clusters are formed. The
$\nabla \cdot \vec{v}$ would be negative at these regions due to the
inflow of mass towards them. Combining these information in
\autoref{eq:two}, we find that the structural elements like sheets,
filaments and clusters would act as a sink of configuration
entropy. The configuration entropy dissipates through these structural
elements of the cosmic web. On the other hand, although the density
decreases in the underdense regions but the $\nabla \cdot \vec{v}$
remains a positive quantity due to the mass outflow from these
regions. The \autoref{eq:two} under such a situation would always lead
to an increase in the configuration entropy. So the underdense regions
or the voids can be regarded as the source which generates
configuration entropy.

Many earlier studies \citep{kauffmann,elad,hoyle,plionis,platen}
indicate that voids in the galaxy distribution occupy $\sim 95\%$ of
the total volume. \citet{colberg} find that the void volume fraction
in a set of GIF2 simulations of the $\Lambda$CDM model increases from
$2.7\%$ at redshift 3 to $61.2\%$ at redshift 0. A recent analysis of
the Millennium simulations \citep{springel, boylan} by \citet{cautun}
using the NEXUS algorithm finds that at present the voids occupy the
largest volume fraction ($\sim 77\%$) in the Universe but contain only
$\sim 15\%$ of the total mass content. Their findings show that the
filaments host the greatest share of mass ($\sim 50\%$) occupying only
$\sim 6\%$ of the volume in the Universe. The sheets occupy $\sim
18\%$ volume and $\sim 24\%$ of the mass content. The nodes or the
clusters occupy the least amount of volume ($\sim 0.1\%$) while
hosting $\sim 11\%$ of the mass content in the Universe. These
statistics imply that the overdensities like sheets, filaments and the
clusters together host the majority of the mass content ($\sim 85\%$)
of the Universe while occupying only $\sim 33\%$ of the volume of the
Universe.  The $\rho(\vec{x},t) dV$ term in the \autoref{eq:two}
represents the mass content inside each of the volume element
$dV$. The \autoref{eq:two} suggests that a negative divergence of the
peculiar velocity field in the overdense structures like sheets,
filaments and clusters would result in a large decrease in the
configuration entropy of the mass distribution.

The underdense regions or the voids occupy most of the volumes but
contain little amount of mass. The density inside the voids decreases
as matters stream out of them and accumulate around their
periphery. The density within the voids gradually increases outward
from their centre. Consequently, the matter in the void centre moves
outward faster than matter near their periphery. The faster evacuation
from the central regions of the voids leads to a uniform low density
region at their interior \citep{goldberg} which gradually evolve
towards $\delta=-1$. The divergence of the peculiar velocity field
$\nabla \cdot \vec{v}$ always remains positive inside the underdense
regions or the voids. Initially the configuration entropy would
increase faster inside the voids and then slow down when the matter
evacuation from the central region would turn it into a uniform low
density region.

The configuration entropy rate $\frac{dS_{c}}{dt}$ is negative at the
overdense regions like sheets, filaments and clusters whereas it
remains positive at the underdense regions or the voids. The decrease
and the increase of the configuration entropy from the different parts
of the cosmic web continues with time. At any given instant of time,
the increase in the configuration entropy at the voids plus the
increase in the differential entropy of the velocity field should be
larger than the overall decrease of the configuration entropy at the
sheets, filaments and clusters. This is particularly true in the
absence of any other major source of entropy. We expect the Universe
as a whole to behave like a thermodynamical system and the total
entropy of the Universe must always increase with time.

When structure formation in the Universe enters the non-linear regime,
the growth of the non-linear structures and the emergence of the
cosmic web would accelerate the dissipation of the configuration
entropy. The Universe soon reaches a stage when the dissipation rate
of the configuration entropy from the overdense regions is about to
overcome its growth rate from the underdense regions and the cosmic
velocity field. The situation can be reversed only if the nonlinear
structures are dissolved or dispersed uniformly or the underdense
regions or the voids are uniformly stretched at an accelerated
rate. The non-linear structures can not be dispersed due the presence
of their gravity and hence can not be a viable option. However due to
their repulsive and outward peculiar gravitational acceleration, the
voids naturally expand faster than the Hubble flow. It may be noted
that the voids are very low density regions where the divergence of
the linear peculiar velocity field $\nabla \cdot \vec{v} \propto -
\frac{\del \delta(\vec{x},t)}{\del t}$ would be a small positive
quantity resulting in a slower increase in the configuration entropy
despite the huge volume occupied by them. Interestingly, if the voids
undergo an accelerated expansion it would lead to a larger $\nabla
\cdot \vec{v}$ inside them due to a larger magnitude of $\frac{\del
  \delta(\vec{x},t)}{\del t}$. It should be noted that this does not
happen due to the normal evacuation of matter from the voids due to
the gravitational field of the non-linear structures at their
periphery. The increase in the divergence of the peculiar velocity
field at the voids purely arises due to a response of the Universe to
the dissipation of the configuration entropy at the overdense
non-linear structures. Incidentally, the accelerated expansion of the
Universe started in the near past when the non-linear structure
formation leads to the emergence of the cosmic web. The accelerated
expansion of the underdense regions or the voids suppress any further
growth of structures on the linear scales. So the freeze out of the
growth of structures on linear scales in the near past may be a
consequence of this accelerated expansion of the voids.

Finally, we relate the deceleration parameter with the configuration
entropy in \autoref{eq:eleven}. The result suggests that we need to
track the evolution of the derivative of the entropy rate to predict
the evolution of the deceleration parameter. In the presence of a
cosmological constant, the deceleration parameter
$q=\frac{\Omega_m}{2}-\Omega_\Lambda$ \citep{sahni2000}. The
deceleration parameter reduces to $q=\frac{3}{2}\Omega_m-1$ in a
critical density universe . This suggest that the value of q would be
$-1$ in a $\Lambda$ dominated Universe. It is interesting to note that
the entropy rate $\frac{dS_c}{dt}=0$ after $S_c$ reaches a constant
value in the $\Lambda$ dominated universe \citep{pandey1}. In this
case, the derivative of the entropy rate $\frac{d^2S_c}{dt^2}$ would
be also zero as the configuration entropy converges to a constant. The
value of $\frac{dP(a)}{da}=0$ in a $\Lambda$ dominated universe as the
growing mode $D(a)=constant$ in this case. So the \autoref{eq:eleven}
also tells us that the deceleration parameter would be $-1$ in a
$\Lambda$ dominated Universe.

In the present work, we propose a physical mechanism which may lead to
the observed accelerated expansion of the Universe. Admittedly, this
does not rule out the possibility of the existence of the dark energy
but provides an alternative which may explain the accelerated
expansion of the Universe without requiring any such fiducial
component.

\section{ACKNOWLEDGEMENT}
The author thanks Diego Pavon for some valuable comments and
suggestions and acknowledges Rishi Khatri for some stimulating
discussions. The author would like to acknowledge financial support
from the SERB, DST, Government of India through the project
EMR/2015/001037. The author would also like to acknowledge IUCAA, Pune
and CTS, IIT, Kharagpur for providing support through associateship
and visitors programme respectively.

\bsp	
\label{lastpage}

\begin{thebibliography}{99}
\bibitem[Amendola \& Tsujikawa(2010)]{de2010} Amendola, L. \& Tsujikawa, S.\ 2010
  Dark Energy: Theory and Observation, Cambridge University Press

\bibitem[Armendariz-Picon et al.(2001)]{armendariz} Armendariz-Picon,
  C., Mukhanov, V., \& Steinhardt, P.~J.\ 2001, \prd, 63, 103510
  
\bibitem[Blumenthal et al.(1984)]{blue} Blumenthal, G.~R., Faber,
  S.~M., Primack, J.~R., \& Rees, M.~J.\ 1984, \nat, 311, 517

\bibitem[Bond et al.(1982)]{bond} Bond, J.~R., Szalay, A.~S., \&
  Turner, M.~S.\ 1982, Physical Review Letters, 48, 1636

\bibitem[Bond et al.(1996)]{bond1} Bond J.~R., Kofman L., Pogosyan
 D.\ 1996, \nat, 380, 603

\bibitem[Boylan-Kolchin et al.(2009)]{boylan} Boylan-Kolchin, M.,
  Springel, V., White, S.~D.~M., Jenkins, A., \& Lemson, G.\ 2009,
  \mnras, 398, 1150

\bibitem[Brans \& Dicke(1961)]{bransdicke} Brans, C. \& Dicke,
  R.~H.\ 1961, Physical Review, 124, 925

\bibitem[Buchdahl(1970)]{buchdahl} Buchdahl, H.~A.\ 1970, \mnras, 150,
  1
\bibitem[Buchert(2000)]{buchert2k} Buchert, T.\ 2000, General
  Relativity and Gravitation, 32, 105

\bibitem[Buchert(1989)]{buchert89} Buchert, T.\ 1989, \aap, 223, 9

\bibitem[Caldwell et al.(1998)]{caldwell} Caldwell, R.~R., Dave, R.,
  \& Steinhardt, P.~J.\ 1998, Physical Review Letters, 80, 1582

\bibitem[Carlberg et al.(1996)]{carlberg} Carlberg, R.~G., Yee,
  H.~K.~C., Ellingson, E., et al.\ 1996, \apj, 462, 32

\bibitem[Cautun et al.(2014)]{cautun} Cautun, M., van de Weygaert, R.,
  Jones, B.~J.~T., \& Frenk, C.~S.\ 2014, \mnras, 441, 2923

\bibitem[Ciecielg et al.(2003)]{ciecielag} Ciecielg, P., Chodorowski,
  M.~J., Kiraga, M., et al.\ 2003, \mnras, 339, 641

\bibitem[Colberg et al.(2005)]{colberg} Colberg, J.~M., Sheth, R.~K.,
  Diaferio, A., Gao, L., \& Yoshida, N.\ 2005, \mnras, 360, 216

\bibitem[Copeland et al.(2006)]{copeland} Copeland, E.~J., Sami, M.,
  \& Tsujikawa, S.\ 2006, International Journal of Modern Physics D,
  15, 1753

\bibitem[Coles et al.(1993)]{coles93} Coles, P., Melott, A.~L., \&
  Shandarin, S.~F.\ 1993, \mnras, 260, 765

\bibitem[Das \& Pandey(2019)]{das} Das, B., \& Pandey, B.\ 2019,
  \mnras, 482, 3219

\bibitem[Davis et al.(1985)]{davis} Davis, M.,
  Efstathiou, G., Frenk, C.~S., \& White, S.~D.~M.\ 1985, \apj, 292,
  371

\bibitem[Doroshkevich(1970)]{doros} Doroshkevich, A.~G.\ 1970
  Astrophysics, 6, 320

\bibitem[Doroshkevich et al.(1980)]{doros80} Doroshkevich, A.~G.,
  Kotok, E.~V., Poliudov, A.~N., et al.\ 1980, \mnras, 192, 321

\bibitem[Easson et al.(2011)]{easson} Easson, D.~A., Frampton, P.~H.,
  \& Smoot, G.~F.\ 2011, Physics Letters B, 696, 273

\bibitem[El-Ad et al.(1996)]{elad} El-Ad, H., Piran, T., \& da Costa,
  L.~N.\ 1996, \apjl, 462, L13

\bibitem[Goldberg \& Vogeley(2004)]{goldberg} Goldberg, D.~M., \&
  Vogeley, M.~S.\ 2004, \apj, 605, 1

\bibitem[Hahn et al.(2015)]{hahn} Hahn, O., Angulo, R.~E., \& Abel,
  T.\ 2015, \mnras, 454, 3920

\bibitem[Hidding et al.(2014)]{hidding14} Hidding, J., Shandarin,
  S.~F., \& van de Weygaert, R.\ 2014, \mnras, 437, 3442

\bibitem[Hoyle \& Vogeley(2002)]{hoyle} Hoyle, F., \& Vogeley,
  M.~S.\ 2002, \apj, 566, 641

\bibitem[Hubble (1929)]{hubble} Hubble, E. \ 1929, Proceedings of the
  National Academy of Sciences of the U.S.A., 15, 168
  
\bibitem[Hunt \& Sarkar(2010)]{hunt} Hunt, P. \& Sarkar, S. \ 2010,
  \mnras, 401, 547

\bibitem[Kauffmann \& Fairall(1991)]{kauffmann} Kauffmann, G., \&
  Fairall, A.~P.\ 1991, \mnras, 248, 313

\bibitem[Komatsu et al.(2011)]{komatsu} Komatsu, E., Smith, K.~M.,
  Dunkley, J., et al.\ 2011, \apjs, 192, 18
  
\bibitem[Milton(2003)]{milton} Milton, K.~A.\ 2003, Gravitation and Cosmology, 9, 66 

\bibitem[Mohr et al.(1999)]{mohr} Mohr, J.~J., Mathiesen, B., \&
  Evrard, A.~E.\ 1999, \apj, 517, 627

\bibitem[Nusser \& Dekel(1993)]{nusser} Nusser, A., \& Dekel,
  A.\ 1993, \apj, 405, 437

\bibitem[Padmanabhan(2017)]{paddy} Padmanabhan, T.\ 2017, Comptes
  Rendus Physique, 18, 275,

\bibitem[Padmanabhan \& Padmanabhan(2017)]{paddyhamsa}
  Padmanabhan, T., \& Padmanabhan, H.\ 2017, Physics Letters B, 773,
  81
\bibitem[Pandey(2017)]{pandey1} Pandey, B.\ 2017, \mnras, 471, L77

\bibitem[Pauls \& Melott(1995)]{pauls} Pauls, J.~L., \& Melott, A.~L.\ 1995, \mnras, 274, 99 

\bibitem[Pav{\'o}n \& Radicella(2013)]{pavon1} Pav{\'o}n, D., \&
  Radicella, N.\ 2013, General Relativity and Gravitation, 45, 63

\bibitem[Peebles(1982)]{peebles} Peebles, P.J.~E.\ 1982, \apj, 263, L1

\bibitem[Perlmutter et al.(1999)]{perlmutter} Perlmutter, S.,
  Aldering, G., Goldhaber, G., et al.\ 1999, \apj, 517, 565

\bibitem[Planck Collaboration et al.(2016)]{planck} Planck
  Collaboration, Ade, P.~A.~R., Aghanim, N., et al.\ 2016, \aap, 594,
  A13

\bibitem[Platen et al.(2007)]{platen} Platen, E., van de Weygaert, R.,
  \& Jones, B.~J.~T.\ 2007, \mnras, 380, 551

\bibitem[Plionis \& Basilakos(2002)]{plionis} Plionis, M., \&
  Basilakos, S.\ 2002, \mnras, 330, 399

\bibitem[Radicella \& Pav{\'o}n(2012)]{radicella} Radicella, N., \&
  Pav{\'o}n, D.\ 2012, General Relativity and Gravitation, 44, 685

\bibitem[Ratra \& Peebles(1988)]{ratra} Ratra, B., \& 
  Peebles, P.~J.~E.\ 1988, \prd, 37, 3406 

\bibitem[Riess et al.(1998)]{riess} Riess, A.~G.,
  Filippenko, A.~V., Challis, P., et al.\ 1998, \aj, 116, 1009

\bibitem[Sahni \& Coles(1995)]{sahni95} Sahni, V., \& Coles, P.\ 1995,
  \physrep, 262, 1

\bibitem[Sahni \& Starobinsky(2000)]{sahni2000} Sahni, V., \&
  Starobinsky, A.\ 2000, International Journal of Modern Physics D, 9,
  373.

\bibitem[Sarkar \& Pandey(2018)]{sarkar18} Sarkar, S., \& Pandey,
  B.\ 2018, arXiv:1812.03661

\bibitem[Seto \& Yokoyama(1998)]{seto} Seto, N., \& Yokoyama,
  J.\ 1998, \apjl, 496, L59

\bibitem[Shandarin \& Zeldovich(1989)]{shandarin89} Shandarin, S.~F.,
  \& Zeldovich, Y.~B.\ 1989, Reviews of Modern Physics, 61, 185

\bibitem[Shannon(1948)]{shannon48} Shannon, C. E. \ 1948, Bell
System Technical Journal, 27, 379-423, 623-656

\bibitem[Springel et al.(2005)]{springel} Springel, V., White,
  S.~D.~M., Jenkins, A., et al.\ 2005, \nat, 435, 629

\bibitem[Tassev \& Zaldarriaga(2012)]{tassev} Tassev, S., \&
  Zaldarriaga, M.\ 2012, \jcap, 4, 013

\bibitem[Tomita(2001)]{tomita01} Tomita, K. \ 2001, \mnras, 326, 287

\bibitem[White(2014)]{white14} White, M.\ 2014, \mnras, 439, 3630 

\bibitem[Yoshisato et al.(2006)]{yoshisato} Yoshisato, A., Morikawa,
  M., Gouda, N., \& Mouri, H.\ 2006, \apj, 637, 555

\bibitem[Zeldovich(1970)]{zeldovich} Zeldovich, Y.~B.\ 1970, \aap, 5, 84 

\end{thebibliography}
\end{document}